\documentclass[pdflatex,sn-mathphys-num]{sn-jnl}

\usepackage{graphicx}%
\usepackage{multirow}%
\usepackage{amsmath,amssymb,amsfonts}%
\usepackage{amsthm}%
\usepackage{mathrsfs}%
\usepackage[title]{appendix}%
\usepackage{xcolor}%
\usepackage{textcomp}%
\usepackage{manyfoot}%
\usepackage{booktabs}%
\usepackage{algorithm}%
\usepackage{algorithmicx}%
\usepackage{algpseudocode}%
\usepackage{listings}%
\usepackage{graphicx}
\usepackage[export]{adjustbox}

\theoremstyle{thmstyleone}%
\theoremstyle{thmstyletwo}%

\theoremstyle{thmstylethree}%

\begin{document}

\title{A Comparative Analysis of Distributed Training Strategies for GPT-2}


\author*[1]{\fnm{Ishan} \sur{Patwardhan}}\email{patwardhanip20.comp@coeptech.ac.in}

\author[1]{\fnm{Shubham} \sur{Gandhi}}\email{shubhammg20.comp@coeptech.ac.in}

\author[1]{\fnm{Om} \sur{Khare}}\email{khareom20.comp@coeptech.ac.in}

\author[1]{\fnm{Amit}
\sur{Joshi}}\email{adj.comp@coeptech.ac.in}

\author[1]{\fnm{Suraj}
\sur{Sawant}}\email{sts.comp@coeptech.ac.in}
\affil*[1]{\orgdiv{Department of Computer Science and Engineering}, \orgname{COEP Technological University}, \orgaddress{\street{Wellesley Road}, \city{Pune}, \postcode{411005}, \state{Maharashtra}, \country{India}}}


\abstract{The rapid advancement in Large Language Models has been met with significant challenges in their training processes, primarily due to their considerable computational and memory demands. This research examines parallelization techniques developed to address these challenges, enabling the efficient and scalable training of Large Language Models. A comprehensive analysis of both data and model parallelism strategies, including Fully Sharded Data Parallelism and Distributed Data Parallel frameworks, is provided to assess methods that facilitate efficient model training. Furthermore, the architectural complexities and training methodologies of the Generative Pre-Trained Transformer-2 model are explored. The application of these strategies is further investigated, which is crucial in managing the substantial computational and memory demands of training sophisticated models. This analysis not only highlights the effectiveness of these parallel training strategies in enhancing training efficiency but also their role in enabling the scalable training of large language models. Drawing on recent research findings, through a comprehensive literature review this research underscores the critical role of parallelization techniques in addressing the computational challenges of training state-of-the-art Large Language Models, thereby contributing to the advancement of training more sophisticated and capable artificial intelligence systems.}




\maketitle

\section{Introduction}\label{sec1}

The development of large language models (LLMs) represents a significant milestone in the field of natural language processing, showcasing remarkable capabilities in understanding, generating, and translating human language. As these models grow in size and complexity, they encounter substantial computational and memory challenges that impede their training and deployment. The scalability and efficiency of training these models have become critical issues, necessitating the exploration of innovative parallelization techniques.

This research delves into the realm of parallel training strategies, specifically examining Fully Sharded Data Parallel (FSDP) and Distributed Data Parallel (DDP) methods. These approaches offer potential pathways to mitigate the computational demands of training LLMs by distributing the workload across multiple processing units. By evaluating these strategies in the context of training the Generative Pre-Trained Transformer-2 model (GPT-2) model, this research aims to uncover insights that can further the efficient and scalable development of LLMs.

The computational hurdles associated with LLM training are not insurmountable. Recent advancements have laid the groundwork for overcoming these challenges through various parallelization strategies \cite{paszke2019pytorch}. For instance, the introduction of FSDP and DDP has marked a significant step towards enhancing the training efficiency of LLMs \cite{rajbhandari2020zero}\cite{shoeybi2019megatron}. These techniques, by distributing the training process across multiple Graphics processing units (GPUs) or nodes, promise to alleviate the bottlenecks of computational resources and memory constraints\cite{huang2019gpipe}.

Moreover, the role of GPU computing has become increasingly pivotal in the training of LLMs\cite{harlap2018pipedream}. The capabilities of GPUs to perform parallel operations have opened new avenues for accelerating the training process, making it feasible to train larger models within reasonable timeframe. This is complemented by innovative techniques such as dense-and-sparse quantization \cite{kim2019parallax}\cite{rhu2018compressing}, which aim to reduce the computational load while preserving the model's performance.

The exploration of hybrid parallelization strategies further illustrates the ongoing efforts to optimize LLM training. These strategies combine the strengths of data and model parallelism to improve training throughput\cite{jia2019beyond}. Additionally, the advancements in dynamic memory management and the development of frameworks like PyTorch's FSDP highlight the continuous evolution of tools and methods to support the efficient training of LLMs\cite{chen2018efficient}.

The following sections of this research are organized as follows: The literature review discusses various techniques for training large language models. It then reviews different training techniques by exploring their architecture and algorithm to implement those techniques. This is followed by the methodology where the framework for evaluating LLMs is stated. The metrics obtained are then used for evaluating different training techniques in the results section. This is followed by conclusion and future scope.

\subsection{Problem Statement}
The advancement of generative artifical intelligence (AI), underscored by the development of LLMs such as GPT-2, has notably enhanced capabilities in natural language processing, including language understanding, generation, and translation. However, the training of these models is challenged by their significant computational and memory demands. To address these challenges, innovative solutions are required to ensure the feasibility and efficiency of their development. The focus on the development and optimization of LLMs has led to the exploration of data parallelization techniques, notably FSDP and DDP. These techniques have been identified as efficient methods for distributing the training workload across multiple GPUs and computational nodes, thereby contributing to enhanced training efficiency and scalability.\cite{9668349}\cite{8805338}.

\subsection{Motivation}
The motivation behind this research stems from the need to address the computational bottlenecks inherent in LLM training. Traditional single-GPU training methods are increasingly inadequate for models of this scale, leading to prohibitive training times and resource demands. This situation calls for advanced parallelization techniques that can distribute the training workload across multiple GPUs and computational nodes, thereby enhancing training efficiency and scalability.

This research aims to explore and compare various distributed training strategies, specifically focusing on FSDP and DDP frameworks, to identify the most effective methods for training LLMs like GPT-2.
\subsection{Objectives}

    The objectives of this research are intricately woven around the core aim of enhancing the training efficiency of LLMs like GPT-2 through the utilization of FSDP and DDP strategies. Central to this research is the analysis of key performance metrics—training time, memory usage, throughput, loss, and grad norm—to discern the impact of these parallelization techniques on the training dynamics. By meticulously evaluating these metrics, this research aspires to unearth the parallelization method that strikes an optimal balance between speed and resource efficiency. Furthermore, as LLMs evolve in size and complexity, understanding the scalability of FSDP and DDP becomes paramount. This research delves into the capacity of these strategies to support larger models without causing detrimental effects on training effectiveness or efficiency. Additionally, this research extends the inquiry to the ramifications of these strategies on the model's performance, with a particular focus on learning capabilities across epochs. This comprehensive analysis aims not only to shed light on the efficiency and scalability of FSDP and DDP but also to guide the selection of the most fitting distributed training strategy, thereby facilitating more effective and accessible LLM training.
    Through a comparative analysis of non-distributed training, FSDP, and DDP, this research serves as a guide in selecting the most appropriate distributed training strategy for LLM training.

\section{Literature Review}\label{sec2}

In LLMs and GPU computing, significant advancements have been observed, primarily aimed at enhancing the efficiency and scalability of models. The implementation of dense-and-sparse quantization methods has been particularly noteworthy, as these techniques have been demonstrated to effectively reduce computational demands while maintaining the integrity of model outputs \cite{kim2024squeezellm, dettmers2022llmint8}. A balance has been achieved between minimizing computational load and preserving high-quality outcomes, which is essential for the deployment of sophisticated LLMs in resource-constrained environments.

The domain has also been enriched by the development of strategies aimed at optimizing the throughput of generative inference using GPUs. Innovative approaches that enable the efficient execution of complex language tasks, even within the limitations of single-GPU setups, have been highlighted \cite{sheng2023flexgen}. The significance of these optimizations in expanding the practical deployment of LLMs and ensuring robust performance despite resource limitations has been underscored \cite{yang2023dynamic, Hong}.

Moreover, the exploration of hybrid parallelization strategies has revealed their potential to significantly improve training throughput. The combination of data and model parallelism within these strategies has been shown to offer a viable solution to accelerate the training of deep neural networks, including LLMs \cite{9586300, Ebrahimi}. These methods have paved the way for the development of more advanced training regimes that are capable of adapting to the specific requirements of the training process.

Additionally, the focus on acceleration methods tailored specifically to the training of convolutional neural networks (CNNs) has provided further insights into the optimization of parallel training processes. The importance of strategic resource allocation and efficient scheduling in reducing training times has been emphasized, with valuable lessons being applicable to the training scenarios of LLMs \cite{yang2023predictive}. The diversity of parallelization strategies for neural network training and their contribution to enhancing training efficiency and scalability has also been examined \cite{yao2023empowering}.

Predictive pipelining techniques have been introduced as an effective solution to the compute-latency trade-off, which is a critical consideration in real-time applications of LLMs. These techniques have been designed to optimize the decoding process, ensuring the timely delivery of high-performance outputs \cite{8916357}. The need for models to be both efficient and contextually aware, particularly in the context of LLM-based machine translation, has been highlighted through the integration of cultural awareness, emphasizing the nuanced challenges of LLM training \cite{yang2023inference}.

Sparse-quantized representations have been investigated, offering a pathway to near-lossless weight compression and significantly reducing the resources needed for storage and computation \cite{wu2023zeroquantfp}. This advancement has been recognized as addressing a critical bottleneck in LLM deployment, enabling improved model performance without a substantial increase in resource requirements.

The challenge of GPU idleness has been addressed through the development of optimized data-loader mechanisms, which have been shown to significantly reduce GPU idleness, thus optimizing the utilization and throughput of computational resources \cite{9860533}. These optimizations are crucial in ensuring consistent GPU engagement and enhancing the overall efficiency of computational operations.

As advancements continue, there has been a shift towards refining the synergies between LLMs and GPU computing. Efforts to develop more adaptable and efficient frameworks for both the training and inference phases of LLMs have been indicative of this trend \cite{dettmers2023spqr}. These efforts are aimed at enhancing the capabilities and applications of LLMs, promising significant contributions to the field.

Despite notable progress, a discernible research gap has been identified in the creation of universal frameworks that facilitate seamless learning transfer across different LLMs. This gap is particularly evident in multi-modal applications that necessitate the integration of textual and visual data, marking an area ripe for future exploration.

In summary, the integration of GPU computing with LLMs, through advanced parallel data loading and quantization techniques, has established a fertile ground for addressing the computational challenges associated with training and deploying these models. The literature emphasizes the need for continued innovation to fully realize the potential of LLMs across a wide range of artificial intelligence applications, acknowledging the commendable advancements made to date.

In the contemporary landscape of high-performance computing (HPC) and deep 
 learning, significant advancements have been propelled by a suite of pioneering research endeavors in the training of distributed deep neural networks (DNNs) on modern GPU-based clusters. The evolution of Accelerated Data-Parallel Distributed DNN Training methodologies has been marked by a paradigm shift towards optimizing data parallelism in GPU environments. This technique, characterized by its strategic implementation of asynchronous communication and refined data sharding protocols, has been demonstrated to effectively address the prevalent bottlenecks encountered in large-scale DNN training endeavors. Consequently, a pathway has been paved for accelerated computational processes and enhanced model performance \cite{10106337}.

Concurrently, the development of an Asynchronous Distributed Proximal Policy Optimization Training Framework has been recognized as another cornerstone, leveraging GPU architectures for the expeditious training of complex models, particularly within the realm of reinforcement learning. This framework has been distinguished by offering a scalable solution that significantly accelerates convergence rates, thereby facilitating more efficient model training processes \cite{10.1007/978-981-16-6372-7_67}.

Moreover, GPU-based machine learning strategies for the detection of botnet attacks have been explored, highlighting the versatility and robustness of GPU-accelerated frameworks in confronting and mitigating cybersecurity threats. A comprehensive approach leveraging the computational prowess of GPUs has been delineated to enhance real-time detection capabilities, significantly diminishing the susceptibility of networks to advanced cyber threats \cite{MOTYLINSKI2022102918}. Similarly, the investigation into machine learning-based autotuning of GPU-accelerated Computational Fluid Dynamics (CFD) codes has unveiled a promising pathway towards optimizing computational efficiency and accuracy in fluid dynamics simulations, underscoring the cross-disciplinary applicability of GPU-accelerated computational techniques \cite{xue2023machine}.

The discourse on dynamic memory management for GPU-based training of deep neural networks has addressed a pivotal challenge within the domain of deep learning—memory bottlenecks. The introduction of innovative memory allocation and recycling strategies has aimed to facilitate the training of larger, more complex models on GPUs, contributing to the scalability and efficiency of deep learning endeavors \cite{8820980}. In parallel, the proposition of OSDP: Optimal Sharded Data Parallel for Distributed Deep Learning has introduced an optimized framework for data parallelism, seeking to ameliorate workload imbalances and minimize communication overhead in distributed deep learning frameworks \cite{jiang2022osdp}.

The extension of these innovations to the training of Graph Neural Networks (GNNs) through scalable data-parallel distributed training techniques has elucidated the unique challenges and opportunities inherent in graph-based learning. This area of research has contributed to the advancement of GNN methodologies and exemplified the broader applicability and potential of distributed training frameworks across various facets of deep learning \cite{9835176}.

The integration of these research findings into the training of LLMs has opened new vistas in addressing the constraints currently faced in LLM training protocols. By harnessing the scalability, efficiency, and flexibility offered by these advanced distributed training frameworks, a tangible opportunity exists to diminish the computational and temporal demands associated with LLM development. This integration has also highlighted existing research gaps, such as the need for tailored optimization strategies to accommodate the distinct architectural and data requisites of LLMs.

In the pursuit of advancing the frontiers of artificial intelligence through LLMs, the research community has increasingly focused on the development and refinement of parallelization strategies to surmount the computational challenges inherent in training such expansive models. FSDP has emerged as a particularly promising approach, characterized by its innovative technique of partitioning model parameters across multiple GPUs to significantly reduce the per-device memory requirement. This method not only facilitates the scaling of model training processes but also enhances the feasibility of executing large-scale models on distributed computing environments \cite{zhao2023pytorch}.

The implementation of FSDP within the PyTorch framework has offered valuable insights into its practical application, revealing both the potential and the complexities of deploying such advanced parallelization techniques in real-world settings. Experiences gleaned from scaling PyTorch models using FSDP have underscored the critical role of efficient communication protocols, particularly through the adoption of compression-assisted all-gather and reduce-scatter operations. These protocols optimize the bandwidth consumption for parameter synchronization, a pivotal factor in maintaining training speed and efficiency across distributed systems \cite{10177384}.

Furthermore, advancements in quantized distributed training have introduced quantization into the distributed training workflow, significantly reducing the computational resources required while providing convergence guarantees. Such guarantees ensure that the model's performance remains robust, addressing one of the primary concerns associated with quantization techniques \cite{markov2023quantized}. The successful training of a model with over a trillion parameters using FSDP on cloud platforms further exemplifies the scalability afforded by FSDP, demonstrating how leveraging cloud computing resources can make the training of previously inconceivable model sizes a reality \cite{trillion}.

The broader ecosystem of parallel and distributed training systems designed to support large models plays a crucial role in the practical application of FSDP. These systems provide the necessary infrastructure to accommodate the unique demands of FSDP, including the management of data flow and parameter synchronization across a distributed network \cite{nagrecha2023systems}. Optimizing multi-GPU parallelization strategies complements the core principles of FSDP by maximizing resource utilization and minimizing training times, thereby addressing the dual challenges of efficiency and scalability in LLM training.

The versatility of FSDP and related parallelization strategies extend beyond traditional neural networks to more specialized architectures, such as spiking neural networks, highlighting the adaptability of these strategies in meeting diverse computational and architectural requirements \cite{9777367}.

As the research community continues to navigate the complexities of training LLMs, the exploration and refinement of FSDP and analogous parallelization strategies represent a critical area of focus. The collective insights from recent studies underscore the significance of these strategies in overcoming the computational barriers to LLM training. Looking forward, the evolution of FSDP, alongside advances in communication protocols, quantization techniques, and system infrastructure, is poised to drive further breakthroughs in the training of LLMs. This ongoing research not only enhances understanding of efficient model training but also paves the way for the development of more sophisticated, capable, and accessible AI systems, heralding a new era in artificial intelligence research and application.

In essence, advancement in GPU computing steers towards more intricate and competent methods for managing large-scale models. The significance of strategies like quantization, parameter sharding, and the development of optimized distributed training systems is underscored. These breakthroughs not only enhance the capabilities of current hardware technologies but also open new avenues for applications and innovations in deep learning and AI domains with hardware constraints.
\section{Review of Training Techniques}
This section of the research includes a comprehensive exploration of contemporary strategies employed in the training of deep learning models, with a focus on the distinct methodologies of Single GPU Training, FSDP, and DDP. Each approach presents unique advantages and challenges, from the simplicity and accessibility of Single GPU Training, conducive to smaller-scale models and rapid prototyping, to the sophisticated memory efficiency and scalability offered by FSDP and DDP for accommodating larger models and datasets. This section delves into the operational intricacies and comparative benefits of these techniques, highlighting their pivotal roles in overcoming the computational and memory constraints inherent in the training of advanced deep learning models, thereby shaping the future of model development and training efficiency

\subsection{Single GPU Training}
Single GPU training is characterized by its simplicity and direct approach, where the entire model and data reside on a single GPU, avoiding the complexities of distribution across multiple devices. This strategy does not involve the partitioning of model parameters or the synchronization of gradients across different GPUs, which are common in parallel training strategies like FSDP and DDP.

In the Single GPU approach, the model undergoes both the forward and backward passes entirely within one GPU, making it straightforward but constrained by the memory and computational capacity of the single GPU in use. This can limit the size of the model and the batch that can be processed, potentially leading to longer training times compared to distributed methods. However, for smaller models or when computational resources are limited, Single GPU training provides an accessible and uncomplicated pathway to model development and training.

The absence of the need for gradient synchronization across devices, as seen in DDP, or the sharding of parameters, as required in FSDP, simplifies the training process. However, it also places the entire computational burden and memory demand on a single GPU, which may necessitate careful management of model complexity and batch sizes to prevent memory overflow and ensure efficient training.

Despite its limitations in scalability and efficiency for larger models, Single GPU training remains a valuable approach, particularly for development, testing, and smaller-scale applications. It allows for rapid prototyping and debugging without the overhead of more complex distributed systems, providing a vital role in the ecosystem of deep learning training strategies.

\subsection{Fully Sharded Data Parallel}
\begin{sidewaysfigure}
\centering
\includegraphics[width=\textwidth]{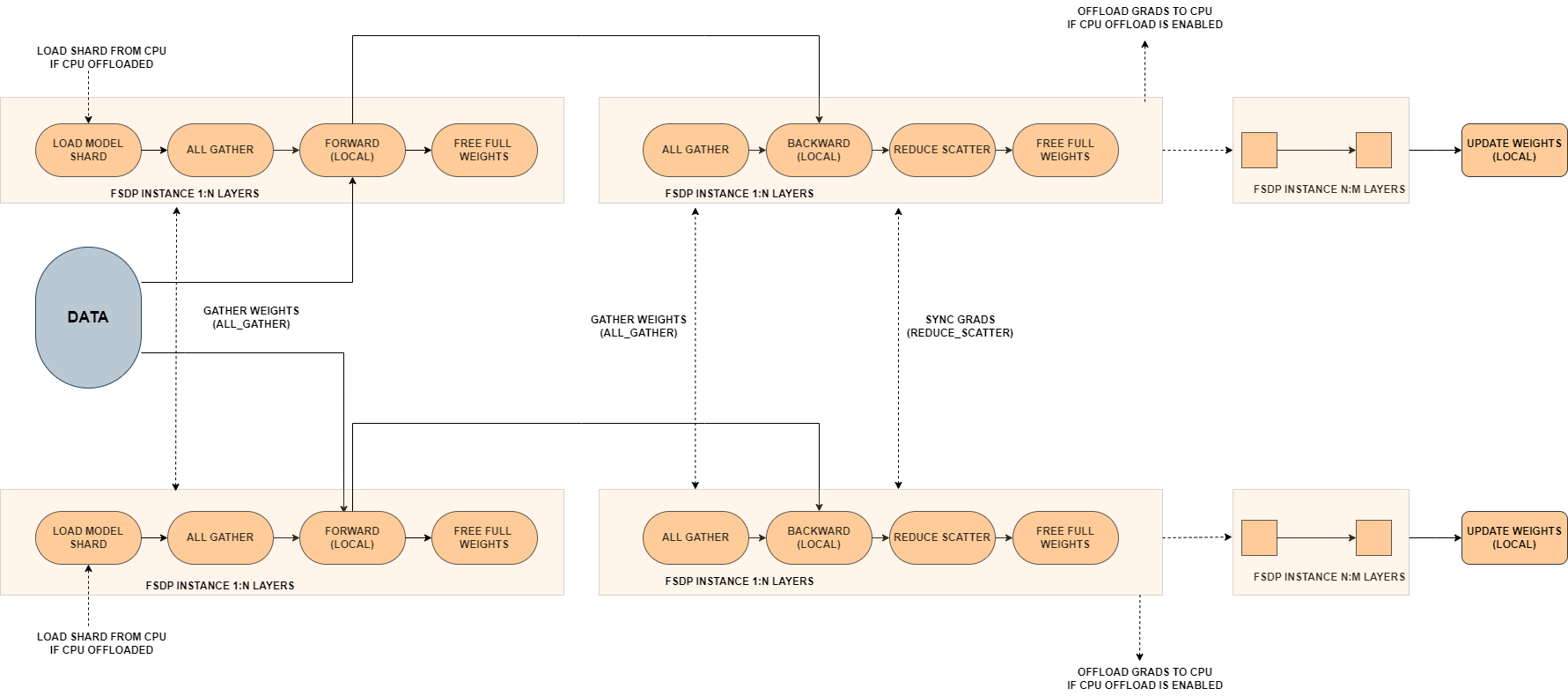}
\caption{Architecture of PyTorch's FSDP}
\label{fig:fsdp}
\end{sidewaysfigure}

FSDP has been designed as a parallel training strategy to address the challenges of memory constraints and to optimize the scalability of deep learning models across multiple GPUs. Challenges are often faced by traditional parallel training methods, such as Data Parallelism, when large models and datasets are involved, primarily due to the memory limitations on individual GPUs. By sharding the model's parameters, gradients, and optimizer states across all available GPUs, FSDP effectively distributes the computational workload and memory usage, as can be seen in the implementation of FSDP class in PyTorch \cite{Paszke_PyTorch_An_Imperative_2019}. The architecture of FSDP is depicted in Figure \ref{fig:fsdp}.


In the FSDP approach, model parameters are divided into shards, each located on a different GPU. Each GPU, during the forward pass, processes a portion of the input data and computes local gradients with respect to its shard of parameters. These local gradients are then synchronized across all GPUs during the backward pass, leading to the update of each shard of parameters with the collective gradients. This ensures the maintenance of consistent model states across all GPUs and the collective learning from the same global gradients.

The significant reduction in the memory footprint on each GPU is one of FSDP's key advantages. The distribution of parameters across multiple GPUs divides the memory requirement for storing model parameters, enabling larger models to be accommodated within the available GPU memory. Additionally, FSDP facilitates the training of models with larger batch sizes, potentially leading to improved convergence and efficiency.

The employment of mixed precision training and gradient accumulation techniques by FSDP further enhances memory efficiency and scalability. Mixed precision training reduces memory usage by utilizing lower precision floating-point formats for certain computations without compromising training accuracy. Gradient accumulation, on the other hand, aggregates gradients over multiple iterations, allowing for larger effective batch sizes without increasing memory consumption \cite{zhao2023survey}.

The benefits offered by FSDP include:

\begin{enumerate}
\item Reduced memory overhead is achieved through the sharding of model parameters across GPUs, minimizing the memory required on each GPU and enabling the training of larger models with larger batch sizes \cite{zhao2023pytorch}.
\item Scalability is enhanced, allowing deep learning models to scale across multiple GPUs and accommodate larger models and datasets without encountering memory limits \cite{zhang2024tinyllama}.
\item Flexibility is provided, allowing FSDP to be applied to a wide range of model architectures without significant modifications to the model code, making it suitable for various deep learning tasks \cite{zhao2023pytorch}.
\end{enumerate}

\subsection{Distributed Data Parallel}
DDP has been widely adopted as a parallel training technique, distributing the training process across multiple GPUs, typically within the same node or across multiple nodes in a distributed computing environment \cite{karakus2021amazon, tang2020communication}. In DDP, a complete replica of the model is maintained on each GPU, with the input data being divided into mini-batches and distributed among the GPUs for parallel processing.

During the forward pass, the forward activations and loss for its mini-batch of data are independently computed by each GPU. These local losses are aggregated across all GPUs, with the resulting global loss being used to compute gradients during the backward pass. Gradients are then synchronized across all GPUs by DDP, and the model parameters are collectively updated using techniques such as AllReduce, as demonstrated by the pseudo algorithm employed in PyTorch's implementation of DDP \cite{li2020pytorch}.


\begin{algorithm}
\caption{DistributedDataParallel}
\begin{algorithmic}[1]
\Require Process rank $r$, bucket size cap $c$, local model \textit{net}
\Function{constructor}{\textit{net}}
    \If{$r=0$}
        \State broadcast \textit{net} states to other processes
    \EndIf
    \State init buckets, allocate parameters to buckets in the reverse order of \textit{net.parameters()}
    \For{$p$ in \textit{net.parameters()}}
        \State $acc \gets p.\text{grad\_accumulator}$
        \State $acc \rightarrow \text{add\_post\_hook}(\text{autograd\_hook})$
    \EndFor
\EndFunction
\Function{forward}{$\textit{inp}$}
    \State $out \gets \textit{net}(\textit{inp})$
    \State traverse autograd graph from $out$ and mark unused parameters as ready
    \State \Return $out$
\EndFunction
\Function{autograd\_hook}{$\text{param\_index}$}
    \State get bucket $b_i$ and bucket offset using $\text{param\_index}$
    \State get parameter $\textit{var}$ using $\text{param\_index}$
    \State $\textit{view} \gets b_i.\text{narrow}(\text{offset}, \textit{var}.\text{size}())$
    \State $\textit{view}.\text{copy\_}(\textit{var}.\text{grad})$
    \If{all grads in $b_i$ are ready}
        \State mark $b_i$ as ready
    \EndIf
    \State launch AllReduce on ready buckets in order
    \If{all buckets are ready}
        \State block waiting for all AllReduce ops
    \EndIf
\EndFunction

\end{algorithmic}
\end{algorithm}

The efficient gradient synchronization mechanism is highlighted as a key feature of DDP. The optimization of gradient communication across GPUs by DDP aims to minimize synchronization overhead, leveraging asynchronous gradient updates and gradient compression techniques. This ensures consistent updates to the model parameters across each GPU, leading to quicker convergence and enhanced training speed.

Ease of use and scalability are also notable attributes of DDP. Its integration into existing training pipelines is facilitated by its straightforward implementation, requiring minimal modifications to the model or training code. Furthermore, DDP's scalability is evident as it accommodates the increasing number of GPUs, enabling deep learning models to utilize growing computational resources for expedited training times.

Challenges may arise with DDP when large models or datasets are involved, as the necessity for each GPU to store a complete replica of the model can result in memory constraints. Additionally, the effectiveness of DDP might be impacted by factors such as communication bandwidth and network latency in distributed computing environments.

In summary, the features offered by DDP include:

\begin{enumerate}
    \item Efficient Gradient Synchronization: The communication of gradients across GPUs is optimized by DDP, diminishing the time dedicated to synchronization and enhancing the overall training speed.
    \item Ease of Use: The implementation and integration of DDP into existing training pipelines are facilitated by its straightforward nature, necessitating minimal alterations.
    \item Scalability: The well-scaling nature of DDP with the increasing number of GPUs allows for quicker training times as computational resources grow.
\end{enumerate}
\section{Proposed Methodology}
The methodology section delineates a comprehensive framework for the empirical evaluation of LLMs training efficiency for advanced parallelization strategies. Anchored by a rigorous experimental setup, this section meticulously outlines the implementation nuances of the GPT-2 model, leveraging state-of-the-art hardware and software configurations to harness the full potential of parallel processing capabilities. Through a detailed exposition of the model architecture, inclusive of embeddings, layer normalization, attention mechanisms, and feed-forward networks. This research establishes a solid foundation for the subsequent exploration of distributed training paradigms.

Focusing the research on the DDP and FSDP strategies, each scrutinized for their efficacy in optimizing training workflows across multiple GPUs. This inquiry is complemented by a juxtaposition with non-distributed, single-GPU training approaches, offering a holistic view of the scalability and efficiency trade-offs inherent in each method. The methodology is further enriched by an in-depth description of the "All the News" dataset. This dataset's diversity and breadth provide a robust testing ground for assessing the adaptability and performance of LLMs across a spectrum of linguistic contexts.

By integrating detailed implementation specifics with a strategic evaluation of training strategies, the methodology section aims to furnish a nuanced understanding of the operational dynamics at play in the optimization of LLM training.
\subsection{Experimental Setup}
The experiments for optimizing LLM training with advanced parallelization strategies were conducted on a system running Linux with Debian x86\_64 GNU/Linux as the operating system. The machine was equipped with 2 Nvidia T4 GPUs for efficient parallel processing. In terms of the central processing unit (CPU), an Intel(R) Xeon(R) CPU @ 2.30GHz with 16 cores and a clock speed of 2300 MHz was utilized. These GPUs were chosen for their parallel processing capabilities, which significantly expedited the training process of the GPT-2 model \cite{huang2022elixir}. The software environment was configured using Python as the primary programming language, with PyTorch serving as the deep learning framework. Additionally, the Transformers library was leveraged for model components and utilities, while Torch Distributed facilitated the implementation of parallel training strategies. 

\subsection{Implementation details}
The GPT-2 \cite{radford2019language} configuration consisted of a 768-dimensional model with a vocabulary size of 50,257 tokens. It included 12 attention heads, each with a 64-dimensional head, a 3072-dimensional feed-forward network, and 12 transformer layers. The model components were as follows:

\begin{enumerate}
    \item Embeddings: Token and positional embeddings were initialized to represent input tokens and their positions within a sequence, respectively. These embeddings were then combined to form the initial input representation for the transformer layers.
    \item Layer Normalization: Implemented to stabilize the inputs to various components of the transformer layers, improving training efficiency and convergence.
    \item Attention Mechanism: The model employed a scaled dot-product attention mechanism with a causal mask to ensure that predictions for a token could only attend to earlier tokens in the sequence, preserving the autoregressive property.
    \item Feed-Forward Network: Each transformer block contained a two-layer MLP with GELU activation, enhancing the model's ability to capture complex relationships in the data.
\end{enumerate}
The GPT-2 model comprises N Transformer decoder blocks, illustrated in the figure below. Each decoder block consists of components such as a multi-head masked attention layer, a multilayer perceptron layer, normalization, and dropout layers. Utilizing residual connections (represented by branching lines to the addition operator), the block can leverage the input from the previous block. The multi-head masked attention layer (right panel) computes attention scores by employing Q, K, and V vectors to capture the sequential relationships within the input sequence.\cite{yang2023fluid}

\begin{figure*}[ht]
\begin{adjustbox}{left}
\centering
\includegraphics[width=1.00\textwidth]{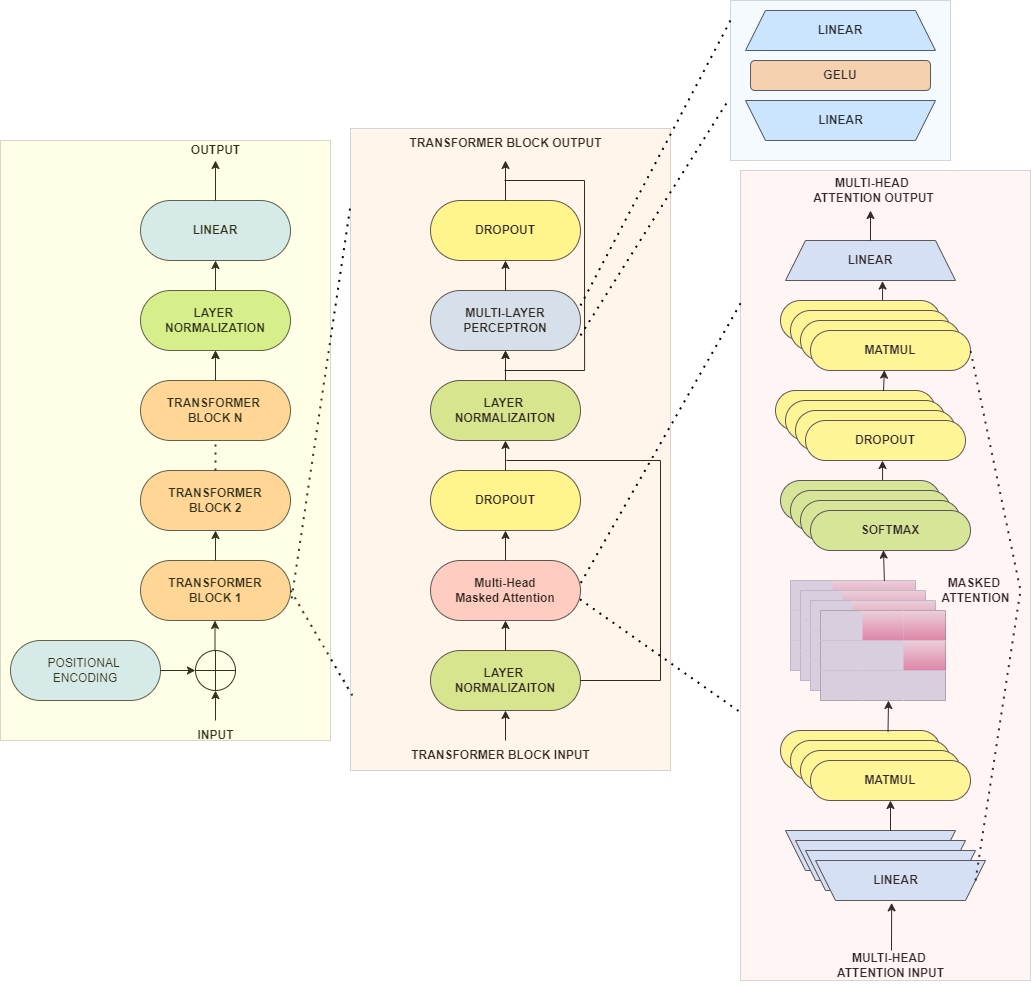}
\end{adjustbox}
\caption{GPT-2 Architecture}
\label{fig:gpt}
\end{figure*}

Illustrated in Figure \ref{fig:gpt}, the input is divided among four separate attention heads. Each head independently processes the input sequence, creating three unique vectors: the query (Q), the key (K), and the value (V). These vectors are then utilized to compute the attention scores using the formula:

\begin{equation}
A(Q, K, V) = \text{softmax}\left(\frac{QK^T}{\sqrt{d_k}}\right)V
\end{equation}

where A represents the matrix of attention scores, Q is the query vector, K transpose is present which
is the transpose of the key vector, V is the value vector, and d indicates the size of the Q and K vectors. The Q vector is associated with the current moment in the input sequence, while K encompasses information across all moments within the sequence. The multiplication of Q and K transpose determines the level of similarity or relationship between different moments in the sequence, allowing the model to focus on the most pertinent sections of the data. The V vector contains the information that the model pays attention to and is employed in creating the output. Following the calculation of attention scores, the various heads are combined through concatenation within a final linear layer, leading to the formation of the overall attention matrix output.

\subsubsection{Training Strategies Evaluated}
\begin{enumerate}
    \item DDP: This strategy involved distributing the model and data across multiple GPUs, synchronizing gradients across all nodes to update model parameters collectively.
    \item FSDP: FSDP was employed to shard model parameters, gradients, and optimizer states across all GPUs, significantly reducing the per-GPU memory footprint and enabling the training of larger models.
    \item Non-Distributed Single GPU Training: This approach utilizes a single T4 GPU to train the model without distributing the workload across multiple devices. Although this method may result in longer training times and constraints on model size due to limited memory capacity, it simplifies the training process and is often used for smaller models or in resource-constrained environments.
\end{enumerate}

\subsubsection{Data Description}
 The "All the News" dataset was used for training the LLMs using different training strategies namely FSDP, and DDP.  The dataset consists of 143,000 news articles from 15 American publications, with a time range primarily between 2016 and July 2017. 

The "All the News" dataset, with its extensive collection of 143,000 news articles from 15 American publications, provides a rich resource for training LLMs due to its diverse content, spanning various topics, writing styles, and perspectives. Training LLMs like GPT models involve several stages, where a dataset like this can be particularly useful. Here's how it can be applied:
\newline
1. Data Preprocessing
Before using the dataset for training, it's crucial to preprocess the text to make it suitable for machine learning algorithms. This involves:
\newline
Cleaning: Removing unnecessary elements like HTML tags, special characters, and formatting.
\newline
Tokenization: Breaking down the text into tokens (e.g., words or subwords) that serve as the basic units for model training.
\newline
Normalization: Standardizing text by lowercasing, stemming, or lemmatization to reduce the variation of tokens.

2. Vocabulary Building
From the preprocessed text, a vocabulary is constructed, listing all unique tokens that appear in the dataset. This vocabulary is crucial for converting text into numerical form that models can understand. For LLMs, dealing with a large vocabulary efficiently is key to handling diverse datasets.
\newline
3. Training Data Preparation
The dataset is divided into segments (e.g., sentences or paragraphs) that are used for training. Each segment is paired with a target outcome, which, in the case of LLMs, is often the next token in the sequence. This trains the model to predict the likelihood of each token given the preceding context.
\newline
4. Model Training
During training, the LLM learns to understand patterns, contexts, and relationships between tokens in the dataset. It involves adjusting the model's parameters to minimize the difference between the predicted and actual outcomes. For a dataset like "All the News," the model learns from a wide variety of linguistic features, including syntax, semantics, and the use of language in different contexts (e.g., political, economic, cultural).
\newline
5. Fine-tuning for Specific Tasks
After the initial training on a broad dataset, LLMs can be fine-tuned on smaller, task-specific datasets. This step adjusts the model to perform well on particular applications, such as sentiment analysis, summarization, or question answering. The diverse content of the "All the News" dataset makes it an excellent resource for initial broad training, providing a foundation that can be refined for various specific tasks.
\newline

The diverse and extensive nature of the "All the News" dataset allows for training LLMs that are robust and versatile, capable of understanding and generating text across a wide range of topics and styles. The dataset's variety in political alignment, publication medium, and subject matter can help mitigate model biases by exposing the model to a wide range of viewpoints and writing styles.

\subsubsection{Performance Metrics}
To assess the effectiveness of the DDP and FSDP training strategies, the following metrics were recorded:

\begin{enumerate}
    \item Training Loss: Monitored to gauge the model's learning progress over epochs.
    \item Memory Usage: Peak memory allocation on the GPUs was tracked to evaluate the memory efficiency of the parallel training strategies.
    \item Training Time: The total time taken to complete the training process was measured, providing insights into the computational efficiency of DDP and FSDP.
    \item {Throughput}: The throughput obtained in the case of different training strategies is compared
    \item {Grad Norm}: The fluctuations in grad norm across training strategies are compared.
\end{enumerate}

\section{Results and Discussion}\label{sec2}
This section presents the results, comparing different training methods over metrics such as training time, throughput, gradient normalization, memory usage, and loss, namely FSDP, DDP, and traditional non-distributed single GPU training approaches. The experiments were conducted on a high-performance computing system under a Linux environment, specifically Debian x86\_64 GNU/Linux. The hardware setup included dual Nvidia T4 GPUs, known for their efficient parallel processing capabilities, alongside an Intel(R) Xeon(R) CPU @ 2.30GHz, boasting 16 cores and a clock speed of 2300 MHz, to ensure robust computational support.

\subsection{Throughput}
Throughput, in the context of GPU training, refers to the amount of data processed per unit of time. It's a critical performance metric, especially in large-scale training tasks, as it directly impacts the efficiency and speed of the learning process. Findings indicate that both DDP and Single GPU configurations maintained a relatively stable throughput across training epochs. Notably, the Single GPU setup consistently achieved a marginally higher throughput compared to DDP, highlighting its efficiency in less complex computational tasks. However, FSDP was observed to have the lowest throughput, suggesting a slower data processing rate which might be attributed to the overhead introduced by data sharding and communication across shards.

\begin{figure*}[ht]
  \centering
  \includegraphics[width=\textwidth]{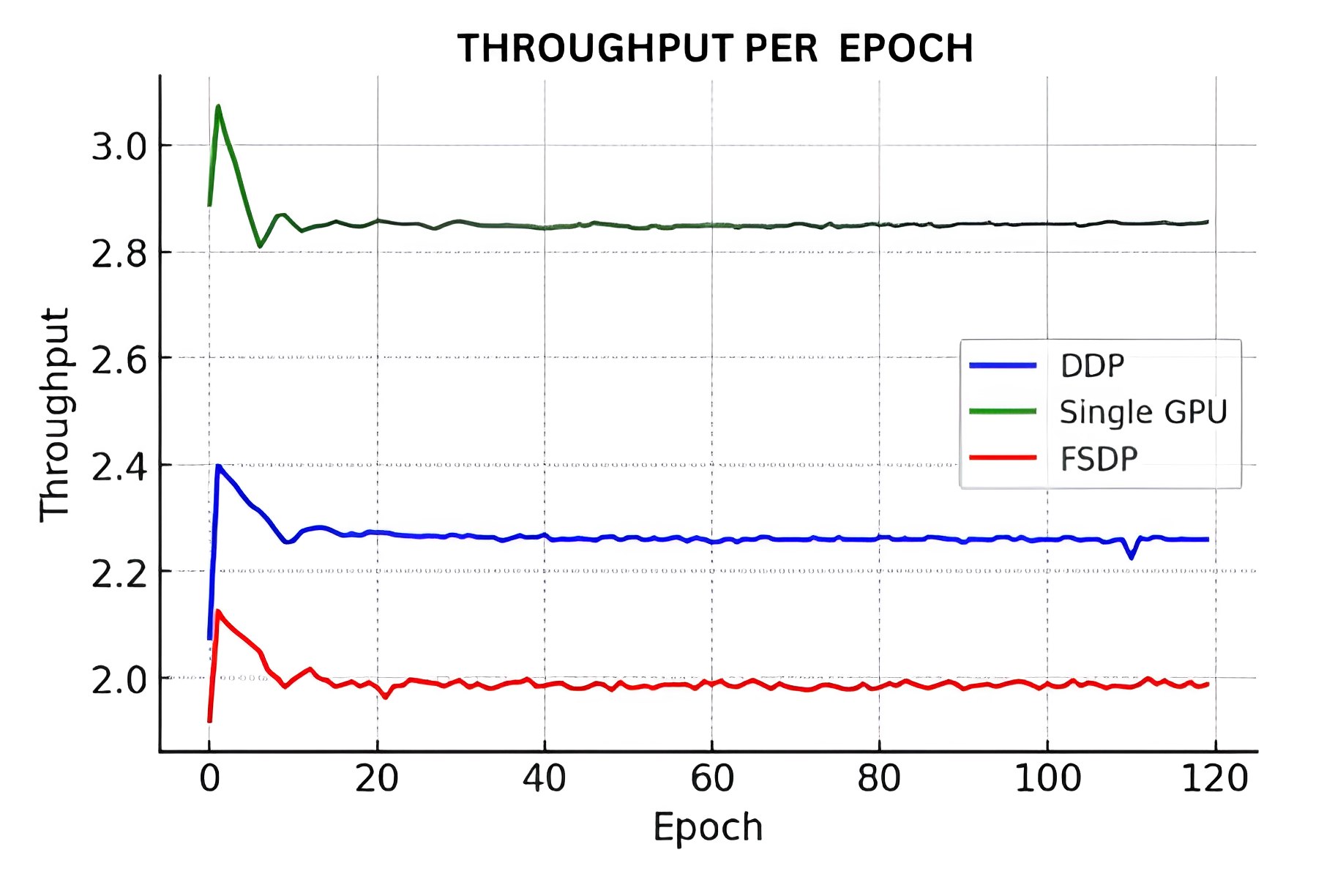}
  \caption{Throughput per Epoch}
  \label{fig:through}
\end{figure*}

Throughput is defined as the amount of data processed per unit of time. Mathematically, it can be expressed as:

\begin{equation}
\text{Throughput} = \frac{\text{Total tokens}}{\text{Training time}}
\end{equation}
In conclusion from Figure \ref{fig:through}:
\begin{itemize}
    \item DDP and Single GPU strategies show relatively stable throughput across epochs, with Single GPU generally achieving higher throughput compared to DDP.
    \item FSDP demonstrates the lowest throughput, suggesting it processes data slower than the other two strategies.
\end{itemize}

\subsection{Gradient Normalization }
Gradient normalization is a technique used to stabilize the training process by scaling the gradients to a manageable range, thereby preventing issues like gradient explosion or vanishing gradients. Analysis showed that all training strategies experienced fluctuations in gradient norms across epochs, which is indicative of variability in training stability. The Single GPU approach, in particular, displayed more significant peaks in gradient norms, hinting at a less stable gradient update process compared to the more distributed approaches like DDP and FSDP. This finding underscores the potential benefits of distributed training methods in maintaining training stability.

\begin{figure*}[ht]
  \centering
  \includegraphics[width=0.99\textwidth]{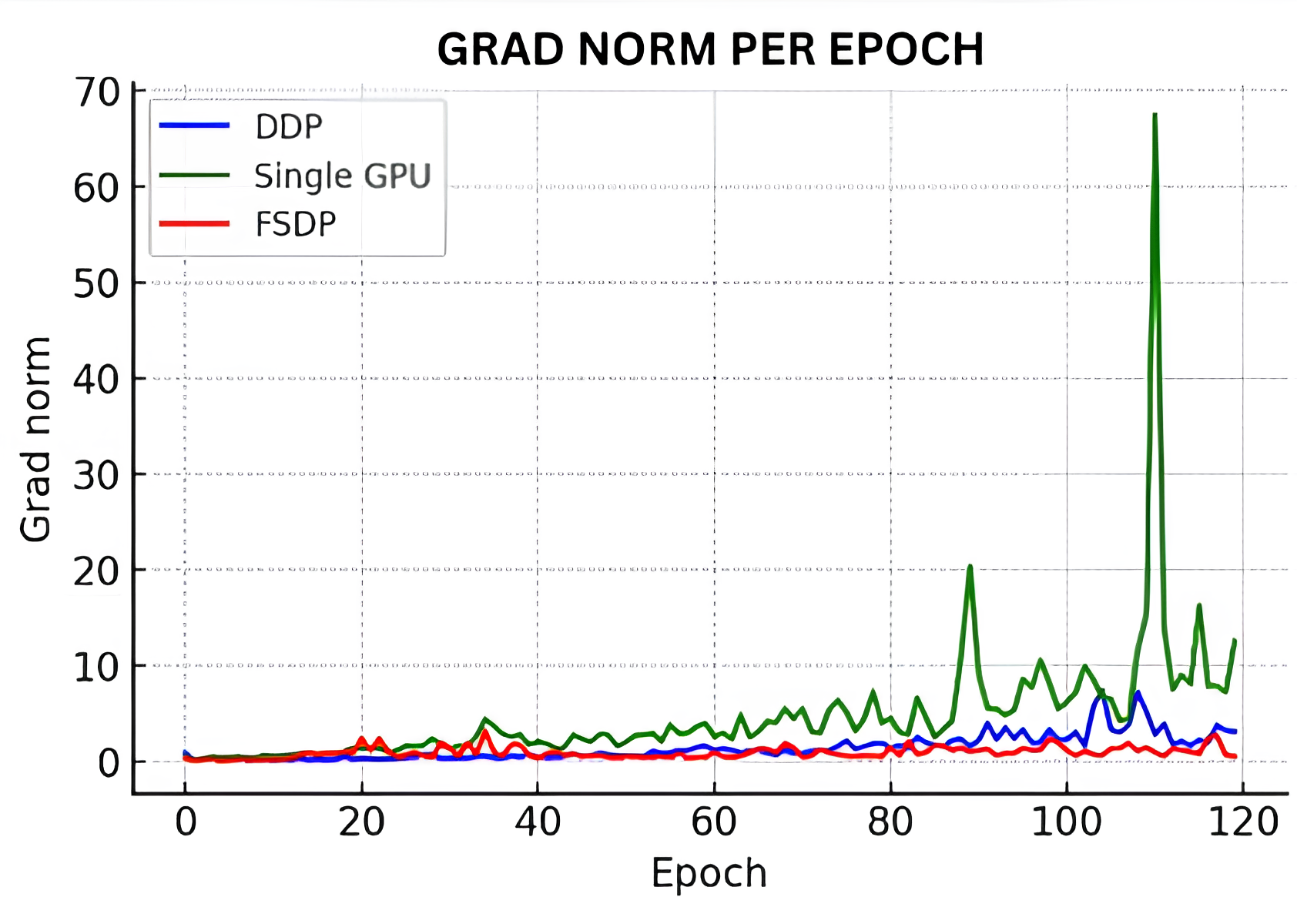}
  \caption{Grad Norm per Epoch}
  \label{fig:grad}
\end{figure*}

\begin{equation}
\nabla L_{\text{normalized}} = \frac{\nabla L}{\|\nabla L\|_2}
\end{equation}

where $\nabla L$ represents the gradients of the loss function $L$ with respect to the model parameters.

In conclusion from Figure \ref{fig:grad}:
\begin{itemize}
    \item All strategies exhibit fluctuations in gradient norm across epochs, indicating variations in training stability.
    \item Single GPU shows higher peaks, suggesting less stability in gradient updates compared to DDP and FSDP.
\end{itemize}

\subsection{Training Time}
Training time per epoch is a straightforward yet vital metric, reflecting the overall efficiency of the training process. The Single GPU method exhibited the longest training times per epoch, suggesting it is less efficient for complex computational tasks compared to its distributed counterparts. On the other hand, DDP and FSDP showed more competitive training times, with FSDP occasionally outperforming DDP. This can be attributed to FSDP's sharded approach, which, despite its lower throughput, can lead to time savings by optimizing memory usage and reducing inter-node communication overhead. 
\begin{figure*}[ht]
  \centering
  \includegraphics[width=0.99\textwidth]{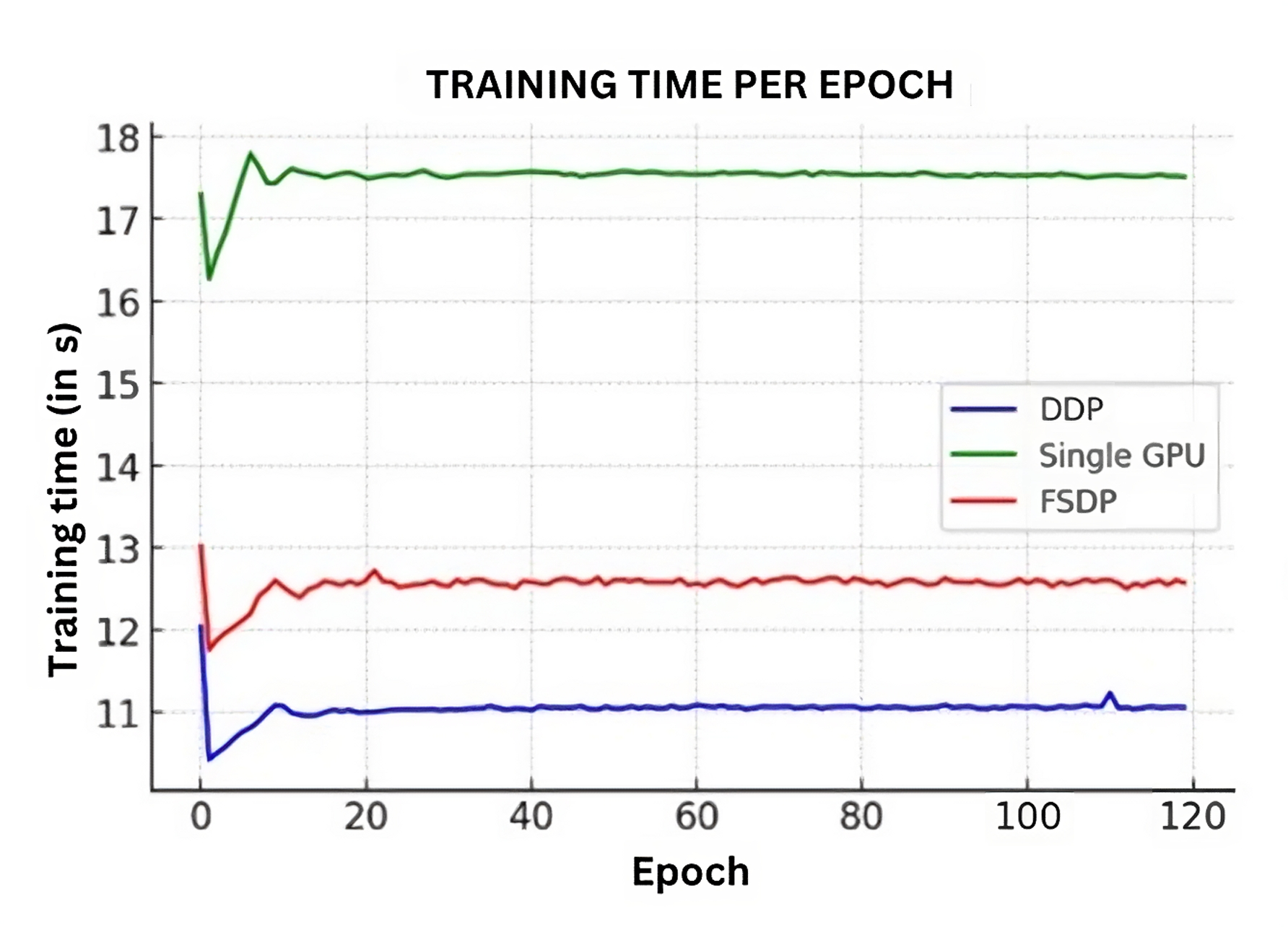}
  \caption{Training Time per Epoch}
  \label{fig:time}
\end{figure*}
In conclusion from Figure \ref{fig:time}:
\begin{itemize}
    \item Single GPU has the longest training time per epoch, indicating it's the slowest strategy among the three.
    \item DDP and FSDP have lower and more comparable training times, with FSDP slightly outperforming DDP in some epochs.
\end{itemize}

\subsection{Memory Usage}
Memory usage is a critical factor in large-scale training tasks, dictating the maximum model size and batch size that can be accommodated. Observations revealed that FSDP demonstrated the most efficient memory usage, maintaining the lowest memory footprint across all epochs. This efficiency is inherent to the sharded data parallel approach, which partitions the model and data across multiple GPUs, thereby minimizing the memory load on individual devices. DDP, while more efficient than Single GPU, still consumed more memory than FSDP. The Single GPU setup, expectedly, had the highest memory consumption due to the lack of workload distribution.

\begin{figure*}[ht]
  \centering
  \includegraphics[width=0.99\textwidth]{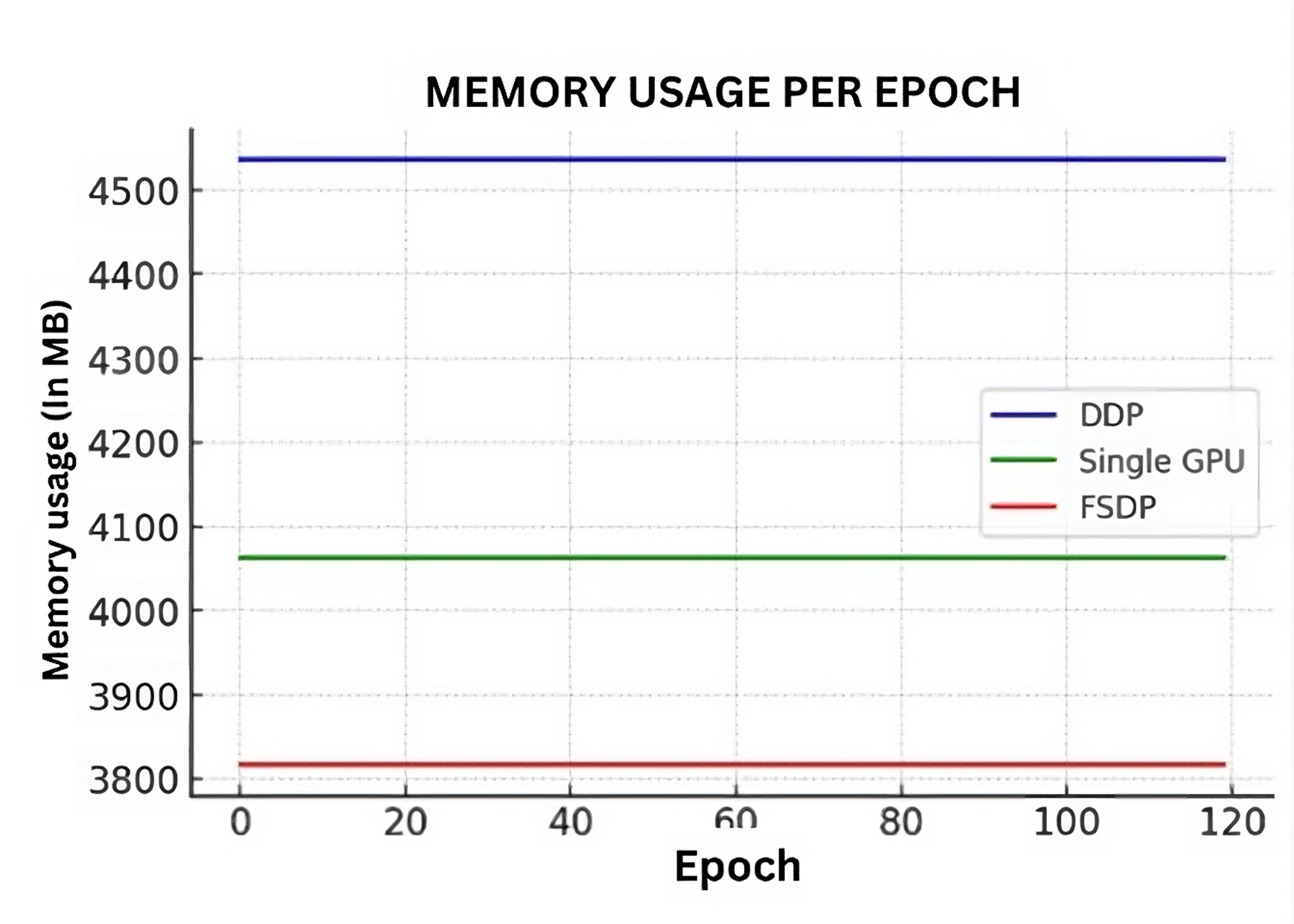}
  \caption{Memory usage per epoch}
  \label{fig:mem}
\end{figure*}
In conclusion from Figure \ref{fig:mem}:
\begin{itemize}
    \item FSDP shows the most efficient memory usage, maintaining the lowest memory footprint across epochs.
    \item DDP uses more memory than FSDP but less than Single GPU.
    \item Single GPU consumes the most memory, which might be due to not distributing the workload across multiple devices.
\end{itemize}

\subsection{Loss}
The reduction in loss over time is a fundamental indicator of the learning efficacy of a training method. Across all evaluated strategies, it is observed that there is a consistent decrease in loss values, confirming that learning is effectively occurring. The initial loss values and the rate at which they decreased were remarkably similar across all strategies. This similarity suggests that, despite the observed differences in throughput, training time, and memory usage, each approach was capable of learning from the data effectively.

\begin{figure*}[ht]
  \centering
  \includegraphics[width=0.99\textwidth]{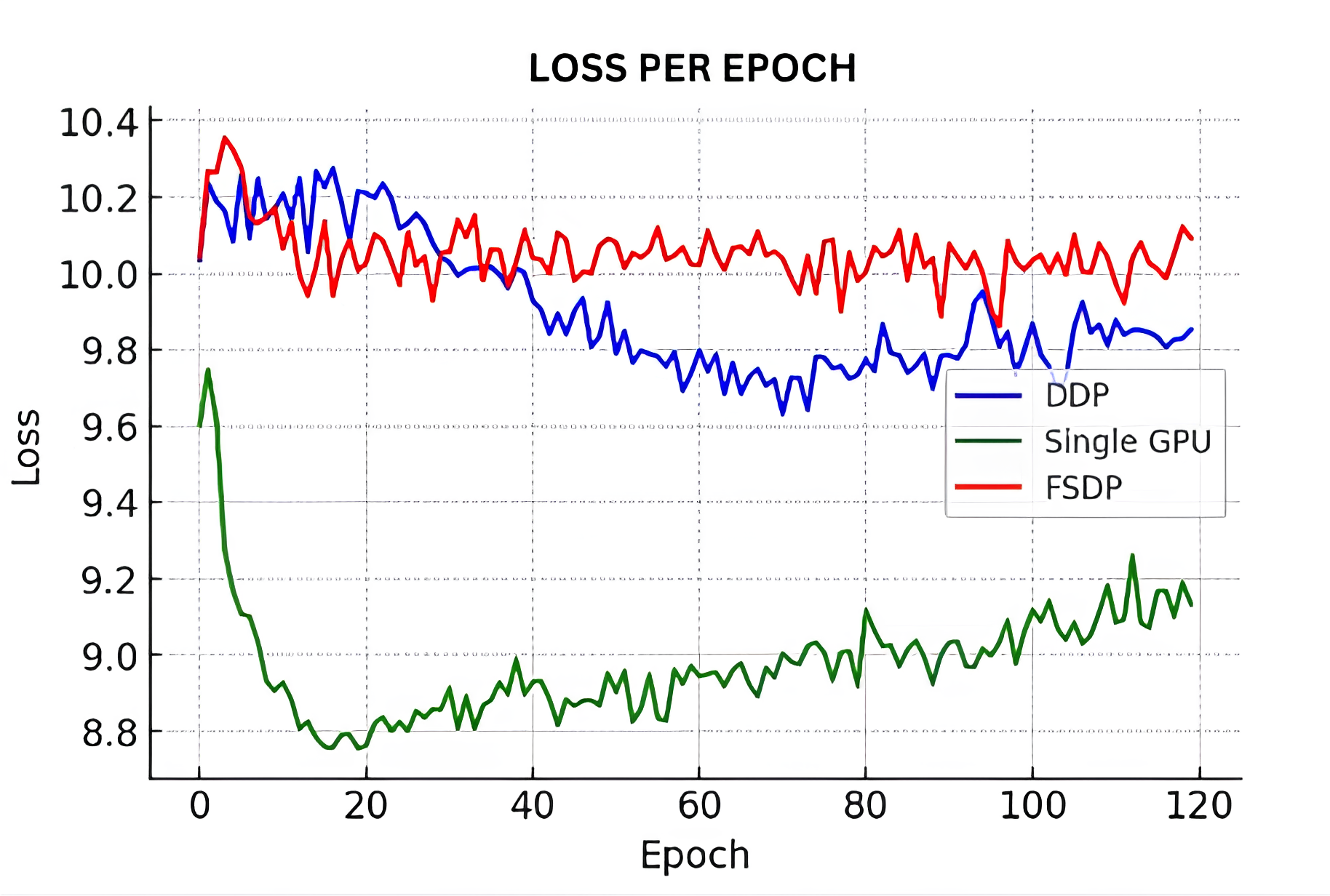}
  \caption{Loss per Epoch}
  \label{fig:loss}
\end{figure*}

The loss is defined as the total loss divided by the total number of tokens. Mathematically, it can be expressed as:
\begin{equation}
\text{Loss} = \frac{\text{Total Loss}}{\text{Total Tokens}}
\end{equation}

where,
\begin{align*}
\text{Total Loss} &= \text{Sum of Losses for Individual Tokens} \\
\text{Total Tokens} &= \text{Number of Tokens in the Dataset}
\end{align*}
In conclusion from Figure \ref{fig:loss}:
\begin{itemize}
    \item Loss decreases over time for all strategies, indicating learning is taking place.
    \item The initial loss values and the rate of decrease are similar across strategies, suggesting that despite differences in throughput, training time, and memory usage, all strategies are effectively learning from the data.
\end{itemize}
 
Comprehensive analysis across pivotal metrics—training time, throughput, gradient normalization, memory usage, and loss—reveals critical insights into the operational efficacy and scalability of these training paradigms.

A key revelation from the research is the nuanced trade-offs between throughput and computational efficiency. While the Single GPU setup demonstrated superior throughput, suggesting an edge in handling less computationally intensive tasks, its elongated training times underscore a diminished efficiency for complex operations. This contrast is stark against the backdrop of FSDP's lower throughput, which, despite its slower data processing rate, achieves significant time savings by optimizing memory usage and minimizing communication overhead. Such findings underscore the importance of selecting a training strategy that aligns with the specific computational demands and scale of the task at hand.

Gradient normalization emerged as a pivotal factor in maintaining training stability, with distributed strategies like DDP and FSDP showcasing a more consistent gradient update process compared to the Single GPU approach. This stability is crucial for preventing issues like gradient explosion or vanishing gradients, which can derail the training process. The distributed methods' ability to stabilize training underscores their potential in handling larger, more complex models where training stability becomes increasingly challenging.

Memory usage, a critical determinant of model and batch size capabilities, was most efficiently managed by FSDP. Its sharded data parallel approach significantly reduces the memory footprint on individual GPUs, enabling the accommodation of larger models and batch sizes. This efficiency is particularly relevant in the era of ever-expanding model sizes, where memory constraints pose a significant bottleneck to scalability and innovation.

Despite the observed disparities from Table \ref{fig:eval} in throughput, training time, and memory usage, all strategies demonstrated a consistent decrease in loss values, affirming their effectiveness in facilitating learning. This consistency suggests that the choice of training strategy, while impactful on efficiency and scalability, does not compromise the model's learning capability. Each approach, with its unique set of advantages and limitations, offers a viable pathway to training LLMs, contingent on the specific requirements of the task.

\begin{table}[h!]
\caption{Comparison of DDP, FSDP, and Single GPU Training}
\centering
\begin{tabular}{lcccc}
\hline
\textbf{Metric} & \textbf{DDP} & \textbf{FSDP} & \textbf{Single GPU} \\
\hline
Avg. Loss & 9.907 & 10.05 & 8.979 \\
Avg. Gradient Norm & 1.404 & 0.919 & 4.664 \\
Total Training Time (seconds) & 1324.003 & 1655.907 & 2101.944 \\
Avg. Memory Usage (MB) & 4536.33 & 3816.28 & 4063.33 \\
Avg. Throughput & 2.266 & 1.992 & 2.854 \\
\hline
\label{fig:eval}
\end{tabular}
\end{table}

\section{Conclusion and Future scope}
This research compares the efficacy of FSDP, DDP, and Single GPU methods for training LLMs. This research examines training time, throughput, gradient normalization, memory usage, and loss, identifying trade-offs for each approach.

FSDP was notably efficient in memory usage, DDP balanced throughput and stability, and all strategies showed consistent learning via loss reduction. These insights enhance the understanding of LLM training dynamics and aid future model development and optimization efforts. It initiates further research into optimizing LLM computational resources and strategies, informing model training dynamics and parallelization strategy selection.

This research is a baseline for enhancing LLM training efficiency and scalability, comparing FSDP, DDP, and Single GPU training. Further research could investigate adaptive parallelization for efficient training of larger models and hybrid approaches that integrate FSDP, DDP, and Single GPU methods for improved efficiency.

As distributed training evolves, it will be critical to AI advancements, aiding in the creation of more advanced models. This research contributes to that progress, aiding in the advancement of LLM applications.
\newline
\newline
\hspace{10pt}
\textbf{\large{Declarations}}
\newline
\textbf{Conflict of interest}: The authors do not have
conflict of interest to declare.
\newline
\textbf{Funding}: The authors have not received any
funding support during experimentation of the
work and writing of the manuscript.
\newline
\textbf{Ethics approval}: This study does not violate and
does not involve moral and ethical statement.
\newline
\textbf{Author’s contributions}: All authors con-
tributed to the work conceptualization and devel-
opment of solution. All authors read and approved
the final manuscript.
\newline
\textbf{Consent for publication}: All authors are aware
of the publication of this manuscript and agreed
to its publication.

\bibliography{sn-bibliography}

\end{document}